\documentclass[twoside]{article}
\usepackage{fleqn,espcrc2}
\usepackage{times,mathptm}
\usepackage[dvips]{graphicx}
\newcommand{\beq}{\begin{equation}}
\newcommand{\eeq}{\end{equation}}
\newcommand{\fcaption}{%
\vspace*{-0.3cm}
\caption%
}
\title{Some Insights into the Method of Center Projection%
\thanks{Presented by \v{S}.\ Olejn{\'\i}k.  Supported in part by 
the Slovak Grant Agency for Science (Grant VEGA No.\ 2/4111/97).}}
\author{%
M.\ Faber\address{Institut f{\"u}r Kernphysik, Tech.\ Univ.\
Wien, A--1040 Vienna, Austria}, 
J.\ Greensite\address{Physics and Astronomy Dept., San Francisco State Univ.,
San Francisco, CA~94117, USA},
\v{S}.\ Olejn{\'\i}k\address{Institute of Physics, Slovak Academy 
of Sciences, SK--842 28 Bratislava, Slovakia}, 
D.\ Yamada$^{\mathrm{b}}$}
\begin{document}
%
%
\begin{abstract}
  We present several new results which pertain to the successes of
  center projection in maximal center gauge (MCG).  In particular, we show why
  any center vortex, inserted ``by hand'' into a thermalized lattice
  configuration, will be among the set of vortices  found by the center
  projection procedure. We show that this ``vortex-finding property'' is
  lost when gauge-field configurations are fixed to Landau gauge 
  prior to the maximal center gauge fixing; this fact accounts for the 
  loss of center dominance in the corresponding projected configurations.
  Variants of maximal center (adjoint Landau) gauge are proposed
  which correctly identify relevant center vortices.
\end{abstract}
\maketitle
%
%
\section{CENTER PROJECTION IN MAXIMAL CENTER GAUGE}\label{projection}
In recent years a wealth of evidence has been accumulated on the lattice in 
favour of the center vortex theory of colour confinement.
Our procedure~\cite{DFGGO98} 
for identifying center vortices consists of the following steps:

1.\ Generate thermalized SU(2) lattice gauge field configurations.

2.\ Fix to {\em maximal center gauge\/} by maximizing:
\beq
{\cal R}[U]=\sum_{x,\mu}\;\Bigl| \mbox{Tr}[U_\mu(x)] \Bigr|^2\;.
\eeq
This in fact is {\em adjoint Landau gauge\/}; the above 
condition is equivalent to maximizing
\beq
{\cal R}[U^A]=\sum_{x,\mu}\;\mbox{Tr}[U^A_\mu(x)]\;.
\eeq

3.\ Make {\em center projection\/} by replacing:
\beq
U_\mu(x)\rightarrow Z_\mu(x) \equiv \mbox{signTr}[U_\mu(x)]\;.
\eeq

4.\ Identify excitations ({\em P-vortices\/}) of the resulting
{$Z_2$} lattice configurations.

P-vortices after center projection  in MCG
appear to be correlated with thick center vortices of full 
configurations~\cite{DFGGO98,DFGO97}. 
Their density scales in MCG~\cite{LRT98}. 
Removal of center vortices destroys confinement
and restores chiral symmetry~\cite{dFDE99}.

In the present paper we address the question why the above 
procedure is able to locate center vortices and why it in some cases 
fails, on lattice configurations preconditioned in a special way.
%
%
\section{VORTEX-FINDING PROPERTY}
The simplest condition which a successful method for locating center
vortices has to fulfill is to be able to find vortices inserted into a lattice
configuration ``by hand''.  
This will be called the {\em ``vortex-finding property''\/}.
Does the method described in Section~\ref{projection}
have this property?

An argument for a positive answer is rather simple:
A center vortex is created, in a configuration {$U$}, by making
a discontinuous gauge transformation.  Call the result {$U'$}.  
Apart from the vortex core, the corresponding link variables in
the adjoint representation, {$U^{A}$} and {$U'^A$}, 
are gauge equivalent.  Let {${\cal R}[U^A] = \mbox{max}$} be a
complete gauge-fixing condition (e.g.\ adjoint Landau gauge) on the
adjoint links.  Then (ignoring both Gribov copies and the core region) 
{$U^A$} and {$U'^A$} are mapped into the same gauge-fixed configuration 
{$\tilde{U}^A$}. The original fundamental link configurations
{$U$} and {$U'$} are thus transformed by the gauge-fixing procedure into 
configurations {$\tilde{U},\ \tilde{U}'$} which correspond to the {\em same\/}
{$\tilde{U}^A$}.  This means that {$\tilde{U},\ \tilde{U}'$} can differ only 
by continuous or discontinuous {$Z_2$} gauge transformations, with the
discontinuous transformation corresponding to the inserted center vortex
in {$U'$}. Upon center projection, {$\tilde{U},\ \tilde{U}' 
\rightarrow Z,\ Z'$}, and the projected 
configurations {$Z,\ Z'$} differ by the same discontinuous
{$Z_2$} transformation.  The discontinuity shows up as an additional thin 
center vortex in {$Z'$}, not present in {$Z$}, 
at the location of the vortex inserted by hand.

This vortex-finding property goes a long way towards explaining the
success of maximal center gauge in locating center vortices in thermalized
lattice configurations, and also suggests that there may be an {infinite
class of gauges} with this property.

However, there are two caveats that could invalidate the argument:

1.\ We have neglected the vortex core region, where 
{$U$} and {$U'$} 
differ by more than a (dis)continuous gauge transformation; and 

2.\ Fixing to ${\cal R}[U^A]=\mbox{max}$ is bedeviled by Gribov copies. 

To find out whether these problems destroy the vortex-finding property, 
we have carried out a series of numerical tests. The simplest is the following:

1.\ Take a set of equilibrium SU(2) configurations. 

2.\ From each configuration make three:

I -- the original one;

II -- the original one with 
{$U_4(x,y,z,t)\rightarrow 
(-\mbox{\bfseries1})\times U_4(x,y,z,t)$} 
for {$t=t_0$},
{$x_1\le x\le x_2$} and all 
{$y$, $z$},
i.e.\ with {2 vortices (one lattice spacing
thick) inserted by hand}.
To hide them a bit, a random gauge copy is made of the configuration
with inserted vortices;

III -- a random copy of {I}.

3.\ Measure:
\beq
G(x)=\frac{\sum_{y,z} <P_{I}(x,y,z) P_{II}(x,y,z)>}
     {\sum_{y,z} <P_{I}(x,y,z) P_{III}(x,y,z)>}\;.
\eeq
{$P_i(x,y,z)$} is the Polyakov line measured on the 
configuration {$i=$I, II, or III}.

If the method correctly identifies the inserted vortices, one simply expects
\beq
       G(x) = \left\{ \begin{array}{rl}
            -1 & x \in [x_1,x_2] \cr
             1 & \mbox{otherwise} \end{array} \right.\;.
\eeq

\begin{figure}[t!]
\centerline{\includegraphics[width=1.0\linewidth]{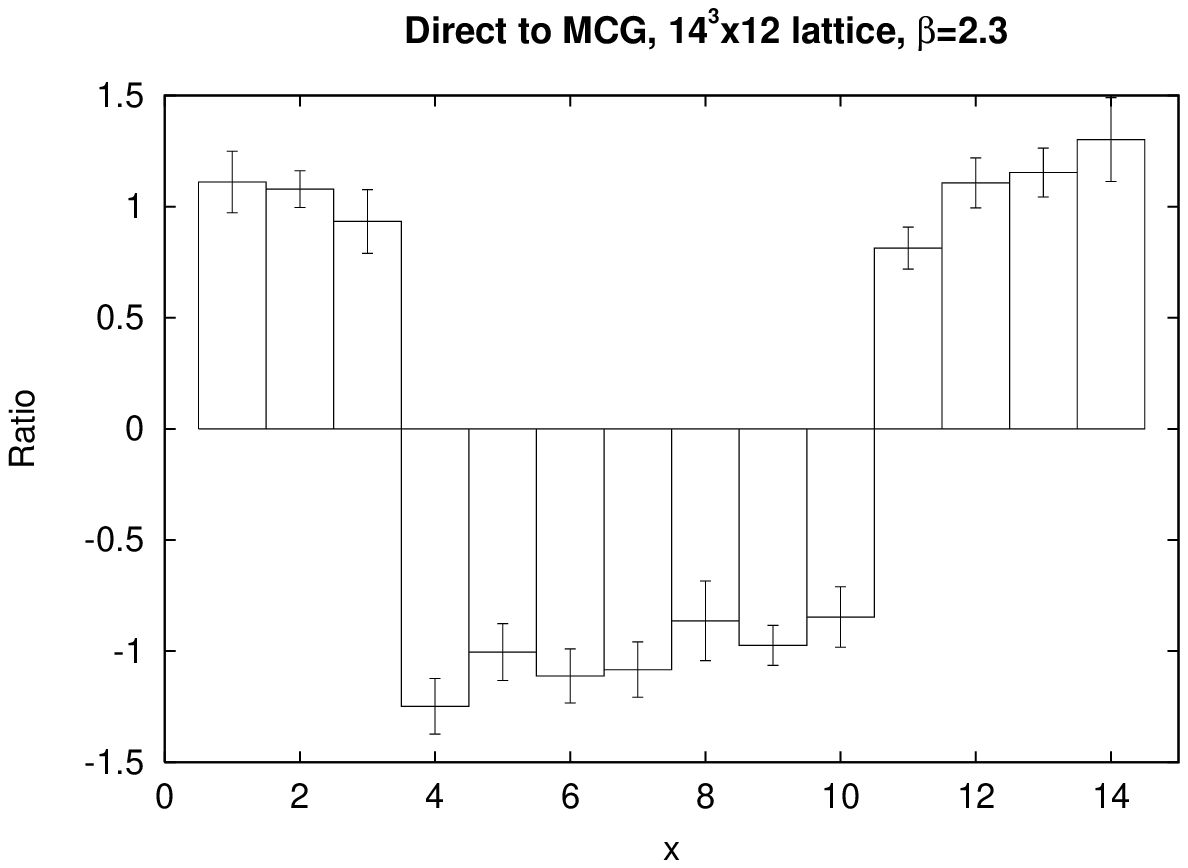}}
\fcaption{Graph of $G(x)$ for configurations
 with thin inserted vortices. Configurations are fixed 
 directly to the maximal center gauge. The discontinuity
 was inserted to the time links within the volume $4\!\leq\!x\!<\!11,\; 
 0\!\leq\!y\!<\!14,\; 0\!\leq\!z\!<\!14$ at the time slice $t=0$.}
\label{fig1}
\end{figure}

The result of the test is shown in Fig. \ref{fig1}. The inserted 
vortices are clearly recognized, and the associated Dirac volume
is found in its correct location.

A more sophisticated test is to insert vortices with a core a few 
lattice spacings thick. Our method also passes that test satisfactorily.
%
%
\section{WHEN GRIBOV COPIES BECOME PROBLEMATIC: 
PRECONDITIONING WITH LANDAU GAUGE}
Gribov copies in maximal center gauge do not seem to be a severe
problem in our procedure; it appears that P-vortex locations vary 
comparatively little, from copy to copy~\cite{DFGGO98}.

However, it has been shown recently~\cite{KT99} that if one first fixes to 
Landau gauge (LG), before relaxation to maximal center gauge, center 
dominance is lost. 

\begin{figure}[t!]
\centerline{\includegraphics[width=1.0\linewidth]{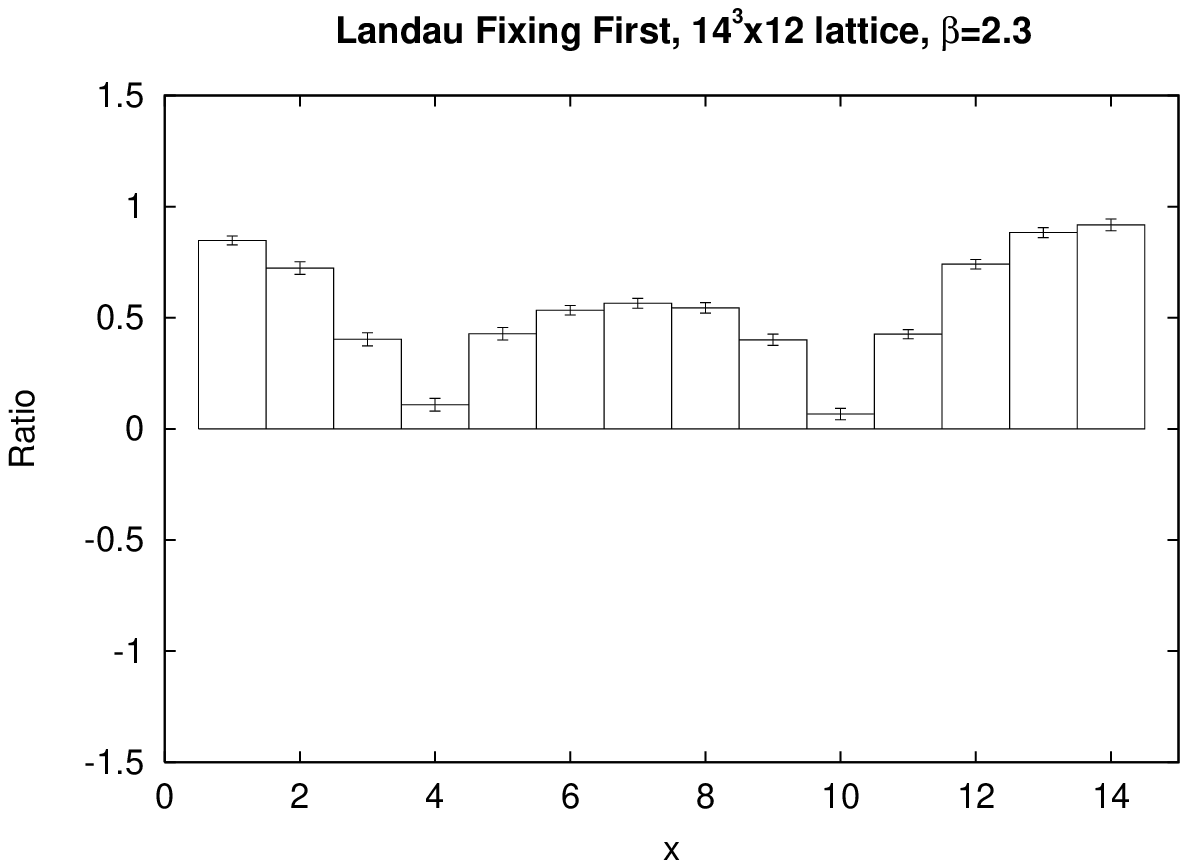}}
\fcaption{Graph of $G(x)$ for configurations
 with thin inserted vortices. Configurations are
 first fixed to the Landau gauge, and only then to MCG.}
\label{fig2}
\end{figure}

This failure has a simple explanation: LG preconditioning 
destroys the vortex-finding property.  This is illustrated by redoing 
the test shown in Fig.\ \ref{fig1}, only with a prior fixing to 
Landau gauge.  The result, shown in Fig.\ \ref{fig2}, is that the
vortex-finding condition is not satisfied; the Dirac
volume is not reliably identified.

The Gribov copy problem, which is fairly harmless on most of the
gauge orbit, seems severe enough to ruin vortex-finding on a tiny
region of the gauge orbit near Landau gauge.%
\addtocounter{footnote}{-1}\footnote{%
Cooling and smoothing, which modify thermalized 
configurations and greatly expand vortex cores, 
also pose some problems 
for center projection.  Whether these are related to the 
Gribov problem, as found in Landau gauge preconditioning, is 
currently under investigation.}
%
%
\section{MCG IS NOT ALONE}
The vortex-finding argument above does not seem to single out MCG.
In fact, there should exist (infinitely) many gauges with the vortex-finding
property. They should fulfill the following
conditions:

1.\ The gauge fixing condition depends on the {\em adjoint\/}
link variable.

2.\ The gauge fixing condition is complete for adjoint links,
leaving a residual {$Z_2$} gauge symmetry for fundamental
links.

3.\ The gauge fixing condition is smooth, the gauge-fixed {adjoint}
link is close to the identity matrix for large $\beta$.

\begin{figure}[t!]
\centerline{\includegraphics[width=1.0\linewidth]{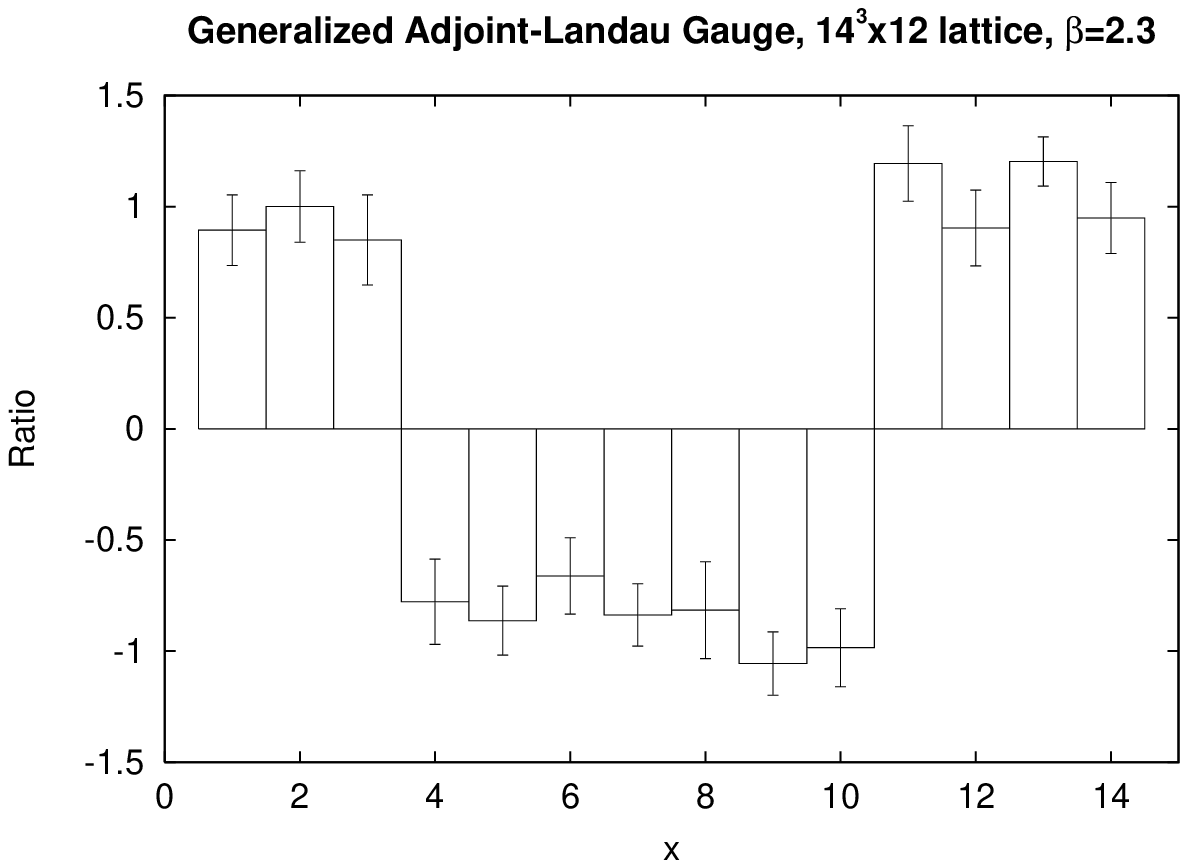}}
\fcaption{Graph of $G(x)$ for configurations
 with thin inserted vortices. Configurations are
 fixed to the gauge maximizing (\ref{genMCG}) with 
 {$c_\mu=\{1,1.5,0.75,1\}$}.}\label{fig3}
\end{figure}

An example is a slight generalization of MCG, namely a gauge
maximizing the quantity
\beq\label{genMCG}
{\cal R}'[U]={\sum_{x,\mu}}\;c_\mu\;\Bigl| \mbox{Tr}[U_\mu(x)] \Bigr|^2\;
\eeq
with some choice of {$c_\mu$}, e.g.\ $c_\mu=\{1,1.5,0.75,1\}$. 
Fig.\ \ref{fig3} shows that this gauge has the vortex-finding property.
Also, center dominance is observed in this gauge, in the same manner as in MCG.

%
%
\section{CONCLUSION} 
We conclude with a sort of tautology: 
{\em To find center vortices, one must use a procedure with
the vortex-finding property.\/}  If that property is destroyed somehow,
e.g.\ by Landau gauge preconditioning, then center vortices are not correctly
i\-den\-tified, and center dominance in the projected configurations is lost.  
This fact does not call into question the physical relevance of 
P-vortices found by our usual method (which {\em has\/} the 
vortex-finding property); that relevance is well-established by 
the strong correlation that 
exists between these objects and gauge-invariant 
observables.

A gauge-fixing technique which complete\-ly avoids the
Gribov copy problem is desirable. A viable alternative has been proposed
by Ph.\ de Forcrand at this conference~\cite{AdFDE99}.

\end{document}